\documentclass[12pt]{article}
\usepackage{graphicx}

\makeatletter \@addtoreset{equation}{section}

\makeatletter\renewcommand\section{\@startsection {section}{1}{\z@}%
                                   {-3.5ex \@plus -1ex \@minus -.2ex}
                                   {2.3ex \@plus.2ex}%
                                   {\normalfont\large\bfseries}}
\renewcommand\subsection{\@startsection{subsection}{2}{\z@}%
                                     {-3.25ex\@plus -1ex \@minus -.2ex}%
                                     {1.5ex \@plus .2ex}%
                                     {\normalfont\bfseries}}

\newcommand{\be}{\begin{equation}}
\newcommand{\ee}{\end{equation}}
\newcommand{\beq}{\begin{eqnarray}}
\newcommand{\eeq}{\end{eqnarray}}

\def\[{\left [}
\def\]{\right ]}
\def\({\left (}
\def\){\right )}

\def\R{{\bf R}}

\def\Z{{\bf Z}}

\def\CN{{\cal N}}

\def\r2{\sqrt{2}}

\def\CF{{\cal F}}

\def\CN{{\cal N}}

\newcommand{\bbibitem}[1]{\bibitem{#1}\marginpar{#1}}

\def\Label#1{\label{#1}%
  \smash{\hbox to0pt{\raise1ex\hbox{\tiny[#1]}\hss}}}
\def\noLabels{\let\Label=\label}
\def\nobbibitem{\let\bbibitem=\bibitem}

\begin{document}
\noLabels 
\nobbibitem 

\begin{titlepage}

\begin{flushright}
{\small
UPR-1154-T  \\ 
LBNL-60486 \\
hep-th/0606118}
\end{flushright}

\vfil\

\begin{center}

{\Large{\bf Four Dimensional Black Hole Microstates: \\ }}
\vspace{3mm} {\Large{\bf From D-branes to Spacetime Foam}} \vfil

\vspace{3mm}

Vijay Balasubramanian\footnote{e-mail:
vijay@physics.upenn.edu}$^{,a}$, Eric G.
Gimon\footnote{e-mail: eggimon@lbl.gov}$^{,b}$$^{,c}$ and Thomas S.
Levi\footnote{e-mail: tslevi@sas.upenn.edu}$^{,a}$
\\

\vspace{8mm}

\bigskip\medskip
\centerline{$^a$ \it David Rittenhouse Laboratories,
  University of Pennsylvania, Philadelphia, PA 19104, USA.}
\smallskip\centerline{$^b$ \it Department of Physics, University of California, Berkeley,
CA 94720, USA.}
\smallskip\centerline{$^c$
\it Theoretical Physics Group, LBNL, Berkeley, CA 94720, USA.}

\vfil

\end{center}
\setcounter{footnote}{0}
\begin{abstract}
\noindent
We propose that every supersymmetric four dimensional black hole of finite area can be split up into microstates made up of primitive half-BPS ``atoms''. The mutual non-locality of the charges of these ``atoms'' binds the state together. In support of this proposal, we display a class of smooth, horizon-free, four dimensional supergravity solutions carrying the charges of black holes, with multiple centers each carrying the charge of a half-BPS state. At vanishing string coupling the solutions collapse to a bound system of intersecting D-branes.  At weak coupling the system expands into the non-compact directions forming a topologically complex geometry.  At strong coupling, a new dimension opens up, and the solutions form a ``foam'' of spheres threaded by flux in M-theory. We propose that this transverse growth of the underlying bound state of constitutent branes is responsible for the emergence of black hole horizons for coarse-grained observables. As such, it suggests the link between the D-brane and ``spacetime foam'' approaches to black hole entropy.
\end{abstract}
\vspace{0.5in}

\end{titlepage}
\renewcommand{\baselinestretch}{1.05}  
\tableofcontents

\newpage

\section{Introduction}

String theory has suggested two different pictures of the
microstates underlying the entropy of black holes.  The first, due
originally to \cite{sen,stromingervafa}, describes the underlying
states as fluctuations of complicated bound states of string
solitons.   These analyses typically apply at very weak coupling,
when there is no macroscopic horizon.   A second picture (see the reviews \cite{mathur1,mathur2})
suggests that, at least in situations with sufficient
supersymmetry, some of the underlying microstates can appear directly in
gravity as a sort of ``spacetime foam''
\cite{mathur1,mathur2,bw2,us,benawarnerfoaming}, the details of which are invisible
to almost all probes \cite{BKS}.\footnote{For 1/2-BPS states in
${\rm AdS}_{5}$ a similar picture has emerged in \cite{LLM,
BdBJS}.  Also see the related comments in \cite{integ,inforecovery}.}     In this picture, the black hole with a horizon is simply the effective semiclassical description of the underlying ``foam''.   The present paper suggests how these two pictures are
connected for four dimensional black holes -- as the string
coupling grows, D-brane bound states that form black hole
microstates grow a transverse size, leading to a gravitational
description as a topologically complex ``spacetime foam''.

We provide evidence for this picture by constructing a large class of smooth,
horizon-free, four dimensional supergravity solutions that have the charges of
macroscopic black holes.  Our construction proceeds by compactifying the
five-dimensional (M-theory on $T^6$) solutions of  \cite{us,bw2}. Typical
geometries contain multiple centers, each carrying the charge of a 1/2-BPS
state. The geometries are characterized by a region of complex topology
containing a ``foam" of spheres threaded by flux.      This leads to a
proposal:  {\it Every supersymmetric four dimensional black hole can be split
up into microstates made of 1/2-BPS ``atoms''.  The mutual nonlocality of the
charges of these ``atoms'' binds the solution together.}

The geometric structures in our solutions scale as the string coupling is changed.
As the string coupling $g_s$ decreases, the region of complex topology shrinks until it is best interpreted in terms of wrapped D-branes whose separations are proportional to
$g_s$.   As $g_s$ decreases further,  the states can be described in the low-energy quiver gauge
theory of a system of intersecting D-branes, following Denef \cite{denefhall}.  Thus we arrive at a picture where quantum gravity microstates associated to a spacetime with fixed asymptotic quantized charges
go through various transitions as the coupling is changed. Every microstate begins life at $g_s=0$ as a ground state of an intersecting D-brane system.   As we increase the coupling, the
microstate makes a transition from a quiver gauge theory in the
Higgs phase, to one in the Coulomb phase.   As we further increase
the coupling, a closed string picture becomes appropriate and, for states having a classical limit, we
obtain the four dimensional solutions described in this
paper.   Still further increasing the coupling to large $g_s$
opens up the eleventh dimension and we find a ``spacetime foam''
in M-theory of two-cycles threaded by flux  \cite{us,bw2}.    A similar flow from D-branes to ``spacetime foam'' has been noted in the topological string \cite{DVVtop}.

\section{Review of five dimensional solutions} \Label{sec-review}

Here we review the candidate smooth, horizon-free microstates for
black holes and black rings in five dimensions that were derived
in \cite{us,bw2}.       We will find candidate microstate
solutions for four dimensional, finite area black holes by
compactifying these geometries.

\paragraph{Basic setup: } M-theory reduced to five dimensions on a 6-torus has 1/8-BPS solutions that only carry membrane charges.  The general ansatz for such solutions was given in \cite{BW} following \cite{reall5d,reallunique}.  The five dimensional non-compact space is written as time fibered over a pseudo-hyperkahler\footnote{The metric is hyperhahler but we allow the signature of HK to flip.  The overall metric remains non-singular and of constant signature because the $Z_i$ will change sign simultaneously.}  base space ($HK$) which we require here to have a
$U(1)$ symmetry.  The metric takes the form:
\beq ds^2 _{11}= -(Z_1 Z_2 Z_3)^{-2/3}( dt + k)^2+(Z_1 Z_2
Z_3)^{1/3} ds^2_{HK}
 + ds^2 _{T^6} , \Label{m-theory metric}
\eeq
where
\be ds^2_{HK} = H^{-1} \sigma^2 + H (dr^2 + r^2 d\theta^2 + r^2
\sin^2\theta
 d\phi^2) \, .
 \Label{HK}
\ee
Here $k$ is a 1-form on the hyperkahler base.   $H$ is a harmonic
function on the flat $\R^3$ parametrized by $(r,\theta,\phi)$ with
poles having integer residues.    Likewise, $\sigma$ is a one-form
on $\R^3$ ($\sigma = d\tau + f_adx^a$ where $\tau$ has period
$4\pi$) satisfying
 \be
\star _3 d\sigma = dH \, .
 \ee
The Hodge dual $\star_3$  acts in the flat $\R^3$ only.   The
metric on the torus is
\beq ds^2_{T^6} &=& (Z_1Z_2 Z_3)^{1/3} \(Z_1^{-1} (dz_1 ^2
+dz_2^2) + Z_2^{-1} (dz_3 ^2 +dz_4^2) + Z_3 ^{-1} (dz_5 ^2
+dz_6^2)\). \eeq
The $Z_i$'s are functions on the hyperkahler base space, and the
associated 2-tori have volumes $V_i$.  The gauge field takes the
form:
\beq
 C_{(3)} &=& - (dt + k) \wedge \( Z_1^{-1}\, dz_1\wedge dz_2 +
Z_2^{-1}\, dz_3\wedge dz_4 + Z_3^{-1}\,
dz_5\wedge dz_6 \) \nonumber \\
&& + 2 \,a^1 \wedge  dz_1\wedge dz_2 + 2\, a^2 \wedge  dz_3\wedge
dz_4 + 2\, a^3 \wedge  dz_5\wedge dz_6 , \Label{cfield} \eeq
where the $a^i$ are one-forms on the base space.  After reduction
on $T^6$ the C-field leads to three separate $U(1)$ bundles, with
connections ${\cal A}^i = -(dt + k)Z^{-1}_i + 2\,a^i$, on the
five-dimensional total space.  Defining $ G^i = da^i,$ \cite{BW}
show that the equations of motion reduce to the three conditions
(here the Hodge operator refers only to the base space $H\!K$):
\be G^i = \star  G^i, \qquad d\!\star \!dZ_i = 2s_{ijk}\, G^j
\wedge G^k, \qquad dk + \star dk = 2\, G^iZ_i, \Label{eq1} \ee
where we define the symmetric tensor $s^{ijk} = |\epsilon^{ijk}|$.

\paragraph{Solution: } This system of equations can be completely solved in terms of seven harmonic
functions in addition to $H$, defined using variables $r_p =
|\vec{x} - \vec{x_p}|$, where $\vec{x}$ is a coordinate in the
$\R^3$ appearing in (\ref{HK}): \be H = \sum_{p\,=1}^{N} {n_p\over
r_p}, \qquad M_i = 1 + \sum_{p\,=1}^{N} {Q^p_i\over 4r_p},\qquad
K = l_0 + \sum_{p\,=1}^{N} {l_p\over r_p},\qquad h^i =
\sum_{p\,=1}^{N} {d_p^i\over 4r_p}. \Label{harmfns} \ee To achieve
an asymptotically flat metric, we require the NUT charge \be n_T =
\sum_p n_p \Label{NUTdef} \ee to equal $1$.  (The standard radial
coordinate for the asymptotic $\R^4$ is  $R = 2\sqrt{r}$.)      At
each point $p$, the $Q_i^p$  measure the membrane charges, $n_p$
sets the Kaluza-Klein monopole charge, $l_p$ measures the angular
momentum associated to the $U(1)$ isometry.  The $d_p^i$ make
contributions to the total 5-brane dipole moment of the solution.
In terms of these functions, the equations of motion can be
solved, giving: \beq Z_i &=& M_i + 2 s_{ijk} h^jh^k/H, \qquad a^i
= (h^i/H) \,\sigma + a^{i}_a dx^a, \qquad d(a_{ia} dx^a) = -\star
_3dh^i, \Label{answer}
\nonumber \\
k &=& k_0 \,\sigma + k_a\, dx^a, \qquad
k_0 = K + 8 H^{-2}\,h^1\,h^2\,h^3 + H^{-1}\,M_i\,h^i \nonumber \\
d(k_adx^a) &=& H \star_3dK - K\,\star_3dH + h^i\, \star_3dM_i -
M_i\,\star_3dh^i \eeq
Requiring the absence of pathologies in these solutions constrains
the parameters in various ways.

\paragraph{Constraints: } The requirement of smoothness (no curvature singularities) constrains the
charges of the harmonic functions:
\be Q^p_i = - s_{ijk} {d^j_pd^k_p\over 2 n_p}, \qquad
 l_p = {d^1_p d^2_p d^3_p\over 16 n_p^2}, \qquad
 l_0 =
  - \sum_i {\sum_p d^i_p \over 4 n_T}
  \equiv - \sum_i {s^i\over 4 n_T}.
 \Label{chargediprel}
\ee It is useful to define \be \tilde{\lambda}_p^i = (d_p^i/n_p -
s^i/n_T),
 \qquad \Gamma_{pq} = { \prod_i (n_p d_q^i - n_q d^i_p)\over n_p^2 n_q^2} \, ,
\ee in terms of which the closure condition for $d(k_adx^a)$,
which makes $k_adx^a$ a globally well defined one-form, becomes
\be
  4 \sum_i n_p \tilde{\lambda}_p^i + \sum_{q=1}^N
{\Gamma_{pq}\over r_{pq}} = 0, \qquad p = 1 ... (N-1). \Label{poleconstraint}
\ee where $r_{pq} = |\vec{x}_p - \vec{x}_q|$.   While (\ref{chargediprel})
constrains the parameters of a smooth solution, (\ref{poleconstraint})
determines the relative separations of the poles in (\ref{harmfns}).  For a fixed
set of charges, there are $N-1$ nonlinear equations relating $3N-3$ variables
(after we fix the center of mass), so if a solution exists for a given set of
charges satisfying (\ref{chargediprel}), the locations of the poles supporting
the charges will typically be a function of $2N-2$ moduli.  We can set three
more free parameters (although these are not completely independent of the
charges) by specifying the angular momentum $J_L$ defined below and its
orientation. Finally, the absence of closed timelike curves and horizons in the
solution requires that
\be (Z_1Z_2Z_3H - k_0^2H^2 - g^{ab}_{\R^3} k_a k_b) > 0
\Label{noCTC} \ee must be satisfied everywhere.   This condition also
guarantees smoothness on the $H = 0$ surface\footnote{At $H=0$ there is an
ergosphere in the solution.  However, unlike a rotating black hole there is no
ergoregion.} and that the metric has constant signature.  It is possible that
(\ref{noCTC}) is implied by (\ref{poleconstraint}) via some kind of gradient
flow argument, but this is not immediately evident.

\paragraph{Charges: }The total membrane charges and angular momenta of the solution are:
\beq
 Q_i &=&  - {1\over 2}\sum_{p\,=1}^{N}
 n_p s_{ijk} \tilde{\lambda}^j_p \tilde{\lambda}^k_p,
 \qquad J_R = \sum_{p\,=1}^{N}n_p \tilde{\lambda}_p^1\tilde{\lambda}_p^2 \tilde{\lambda}_p^3,
 \Label{charges1}  \\
 \qquad J_L &=& 4 | \sum_{p\,=1}^{N}\sum_i n_p \tilde{\lambda}_p^i\vec{x}_p|
             \;=\; {1\over 2}\Big|\sum_{pq} \Gamma_{pq}
        {\vec{x}_p - \vec{x}_q\over |\vec{x}_p - \vec{x}_q|}\Big|  .
\Label{charges2} \eeq The solution  also carries a net 5-brane
dipole moment.   Finally, the absence of Dirac strings in the
$C$-field requires integral quantization conditions:
\beq &&
n_p,\;\; {\pi\Gamma_{pq}\over 4G^{(5)}_N} \in \Z, \qquad d_p^i =
m^i_p/e^i,\qquad
 m^i_p \in \Z,
 \Label{integerquant}
\eeq
with constants
 \be
 e^i = {V_i\over (2\pi)^2 \ell_P^3},\qquad \prod_i (e^i)^{-1} = {4G_5\over \pi}.
\ee Appropriate quantization of the membrane charge and angular
momenta also imposes \be m_p^im_p^j/n_p \in \Z,\qquad
m_p^1m_p^2m_p^3/n_p^2 \in \Z \, . \ee

\section{Reducing to Four Dimensions} \Label{sec-4d}

Type II string theory compactified on a 6-torus has a spectrum of
extremal supersymmetric black holes.  The charge vector of these
black holes transforms in the {\bf 56} representation of the
$E_{7(7)}$ duality group.    The entropy associated with the black hole horizons
is a function of the  quartic
invariant of $E_{7(7)}$ constructed from the charges (see, e.g., \cite{kalloshkol,vijaycargese}).
As a result, to have a finite
entropy, these black holes must have at least four charges.
Furthermore, a generating solution whose $E_{7(7)}$ orbit traces
out the entire 56 dimensional space of extremal black holes must
have at least five charges, of which one pair must be
electric-magnetic duals \cite{5charge,ferraramald}.    We are interested in finding candidate
supergravity microstate solutions for such a generating black
hole.

In the previous section we described a class of candidate microstates for five
dimensional black rings and black holes in M-theory having a $U(1)$ isometry.
These solutions carry three M2-brane charges  and momentum along the $U(1)$
direction.   Upon compactifying along this direction, these charges give rise
to  D2-branes and D0-branes in IIA string theory.     To carry out the
reduction to four dimensions we modify the solution in the previous section so
that the $U(1)$ direction approaches a finite size at infinity.    M5-branes can
wrap on this circle leading to three D4-brane charges in IIA string
theory.   In addition, the now arbitrary NUT charge in the solution leads to
D6-brane charge, giving in total eight charges.\footnote{Given an arbitrary
NUT charge $n_T$, the asymptotic solution has topology $S^3/Z_{n_T}$.  This allows
us to define three $Z_{n_T}$ valued 5-brane monopole charges.  Once we reduce to
 $IIA$ and the M-circle disappears, these give rise to regular integer 4-brane charges;
 shifts of these integer charges by $n_T$ are associated with large gauge transformations of the
B-field.}   A similar procedure, of placing five dimensional solutions in Taub-Nut, has been used to relate five dimensional black holes and rings to four dimensional black holes \cite{4d5d,EEMR}.   All of these charges arise from wrapped D6-branes with fluxes. Since D2s and D4s as well as D0s
and D6s are electric-magnetic duals, these configurations also have the charges
necessary for them to be smooth microstates associated to a finite area,
extremal 1/8-BPS black hole.

\subsection{Introducing Taub-NUT}
\Label{sec-NUT}

The five dimensional solution in Sec.~\ref{sec-review} has a
direction of $U(1)$ isometry.  In order to reduce on this circle
it must approach a finite size at infinity  We can accomplish this
by adding a constant  to $H$ which effectively places the
M2-branes in a Taub-NUT background:
\be H \to H + \delta H ~~~;~~~
\delta H = {4\over L^2}. \ee
The asymptotic circumference of the
circle is $2\pi L$.    We allow the NUT charge (\ref{NUTdef}) to
be arbitrary, the full solution (\ref{m-theory
metric},\ref{cfield}) can then be interpreted in terms of
M2-branes in a Taub-NUT background.

In order to ensure gauge invariance under transformations of the C-field  it is
necessary to also add constants  $\delta h^i$ to the $h^i$ harmonic functions.
In terms of these shifts it is convenient to define
 \be
 \lambda_p^i =
{d^i_p\over n_p} - {4\delta h^i\over \delta H} = {d^i_p\over n_p} -  L^2 \,
\delta h^i,\Label{newlambda}
 \ee
and to re-write the the constant part of $K$ (making sure there are no CTCs at infinity
by bounding the last term) as:
\be
 \delta K =
 {L^4 \over 4} \prod_i \delta h^i - {L^2 \over 4} \sum_i \delta h^i + {L \over 2} \sin\alpha
\Label{deltaK}
 \ee
To achieve a standard locally flat metric at infinity, we set the constant part
of $M_i$ to
 \be \delta M_i = 1 - 2 s_{ijk} {\delta h^j \delta h^k \over
\delta H} .
 \ee
We can also set $k_0 \to 0$ at infinity by setting $\alpha =0$. This
is not necessary for getting a flat asymptotic metric, as we can always set
this constant to zero by shifting the coordinate $\tau$ by a function of $t$.
This adds another layer of unnecessary complication (see Appendix B for the full construction), 
and without loss of generality we will assume that $\alpha =0$. We do require, however,
that the derivative of the angular momentum,
\be
dk= d(k_0 f_adx^a) + d(k_adx^a)
\ee
falls off at infinity. This yields the constraint\footnote{When $\alpha \neq 0$ there is an additional term proportional to
 $\sin\alpha$ on the right hand side of (\ref{extra}).     In the language of D=4, ${\cal N}=2$ supersymmetry, 
this constraint relates the parameter $\alpha$ in (\ref{deltaK}) with the phase of the central charge, in a given gauge, of our 
1/8-BPS solution (see Appendix B). This connection between the angle $\alpha$ and the asymptotic velocity  along the M-direction is 
typical for black rings (see \cite{EEMR}).}
 \be \sum_p \Psi_p = 0,\qquad \Psi_p = \sum_i n_p \lambda^i_p
- {1\over L^2 n_p^2} \prod_i n_p \lambda^i_p \Label{extra} .
 \ee
 In lifting back to five noncompact dimensions the shifts $\delta h^i$ and $\delta H$ are set to
zero so that subleading terms in $h^i$ and $H$ determine the ratio
\be
\lim_{r\to\infty} {4h^i\over H} \to {s^i \over n_T}.
\ee
Thus, in the decompactification limit (\ref{extra}) is tautological.    The
integrability condition on  $k_adx^a$ becomes
\be
  4\Psi_p  + \sum_q {\Gamma_{pq}\over r_{pq}} = 0, \qquad p = 1 ... N.
  \Label{newbubble}
\ee (Compare with (\ref{poleconstraint}) in the $L \to \infty$
decompactification limit.)    Only $N-1$ of these are independent if we take
(\ref{extra}) into account.  Specifying the angular momentum $\vec{J}_L$
provides three more constraints ($J_R$ will be related to the  $D0$-charge):
 \beq
J_R &=& \sum_p n_p \prod_i \lambda^i_p   \\
\vec{J}_L &=& -4\sum_p \Psi_p \,\vec{x}_p= {1 \over 2} \,
\sum_{pq} \Gamma_{pq} \hat{r}_{pq}.
 \eeq
 The second equality for $\vec{J_L}$ used the
constraint equation (\ref{newbubble}).

\subsection{Reduction to IIA}
We obtain a solution to IIA string theory by reducing along
$y=(L/2) \, \tau$  ($y$ has period $2\pi L$) in terms of
dimensionless versions of the eight harmonic functions ($M_i$ is
already dimensionless) \be
    M_0 = - H L^2/4,\qquad K^0 = 4K/L, \qquad K^i = L h^i.
\Label{dimharm} \ee The reduction gives the metric
 \be ds^2_{IIA}
= - J_4^{-1/2} (dt +k_adx^a)^2 + J_4^{1/2} \Big(ds^2_{\R^3} +
\sum_{i=1}^3 (-Z_iM_0)^{-1} ds^2_{T_i}
\Big) \, . \Label{10dmetric} \ee
The radius of the compactification circle is related to the string length and coupling as
\begin{equation}
L = g_s l_s  \Label{Ldef}
\end{equation}
The $T_i$ are flat 2-tori with
volume forms $dV_i$.  The dilaton and form fields are \beq &&
e^{2\Phi} = (J_4)^{3/2} (-Z_1 Z_2 Z_3 M_0^3)^{-1},
\qquad B_2 = -\Big({K^i\over M_0} + {2 k_0\over  L\,Z_i}\Big)dV_i  \Label{dilaton} \\
 &&  C_1 = {L\over 2}f_adx^a - {2M_0^2k_0\over L J_4}(dt + k_a dx^a)\\
 &&  C_3 = \Big[-Z_i^{-1}(dt + k_a dx^a) + 2\vec{a}_i
-\Big({K^i\over M_0} + {2 k_0\over  L\,Z_i}\Big){L\over 2} f_a
dx^a \Big]\wedge dV_i. \eeq $J_4$ is the quartic invariant of
$E_{7(7)}$ constructed from the eight harmonic functions connected
to four electric and magnetic ``charges"  \cite{BKW}
 \beq
J_4  &=& {L^2 \over 4} \left( (Z_1 Z_2 Z_3) H - k_0^2 H^2 \right) \Label{quartic1}\\
&=& M_0K^0(M_iK^i) + M_1K^1(M_2K^2 + M_3K^3) + M_2K^2M_3K^3
\nonumber
 \\
 &-& {1\over 4}(M_\mu K^\mu)^2 - M_0M_1M_2M_3 - K^0K^1K^2K^3
 ,\qquad \mu \in 0\ldots 3
 \Label{quartic}
 \eeq
We have introduced a new radial variable $\rho = 2r/L$ so
that the metric on the $\R^3$ piece takes the standard flat form.

For the further reduction on the $T_i$  to four dimensions it is useful to
define a shifted 3-form field that is invariant under shifts of coordinates in
the $y$ circle and the three 2-tori.
 \be
C_3' =  C_3 - B_2\wedge C_1 = \Big(
{-M_0\over J_4}(2K^ik_0/L - s^{ijk}\, Z_jZ_k/2)(dt + k_adx^a) + 2\vec{a}_i\Big)
\wedge dV_i \, . \ee The quantized $D$-brane charges are \beq
 Q_0^{D6} = {L\over 2}\,\sum_p (-n_p)= {g_s l_s\over 2}\,N_6 , &\quad&
  Q_{i}^{D2}= -{1\over 2L}\,\sum_p s_{ijk}{d_p^jd_p^k\over 2 n_p}
 = -{4G_4\,V_i\over 4\pi^2 g_s l_s^3} N_{i\,2}
 ,  \nonumber \\
 Q^0_{D0} = {1\over 2L^2}\,\sum_p {d_p^1d_p^2d_p^3\over n_p^2}
 = {4G_4\over g_s l_s}\,N_0 ,&\quad&
  Q^i_{D4} = {1\over 2}\,\sum_p d_p^i = 2\pi^2\,{g_sl_s^3\over V_i} N_4^i \, .
  \Label{quantcharge}
 \eeq
In our conventions the D0 and D4-branes are magnetic objects
while the D2 and D6-branes are electric. While the charges defined
here are quantized, they are not invariant under large gauge
transformations of the $B$-field  because of Chern-Simons terms in
the supergravity Lagrangian.  (See \cite{doncharges} for the
difference between quantized  and gauge-invariant charges.)   In
the four dimensional  theory we  will interpret large gauge
transformations of the $B$-field as  $SU(8)$ transformations
inside $E_{7(7)}$.   Our sign conventions are consistent with the
Hodge dual relations $F_6 = *F_4, F_2 = *F_8$.    Each of these
charges is written as a sum over points $p$, and at each point we
can interpret the charges as arising from a D6-brane with fluxes
on it.

\subsection{Reduction to four dimensions and special geometry}
\Label{sec-multi}

\paragraph{$E_7$ structure: }
 Upon reduction to four dimensions, we obtain solutions to ${\cal N} =8$
supergravity~\cite{CremmerJulia}.
This theory has an $E_{7(7)}$ duality group.   The three D2-brane charges and the D6-brane charge transform
in an electric ${\bf 28}$ representation of the maximal compact subroup $SU(8)/\Z_2$, while the three
D4-brane charges and the D0-brane transform in a magnetic ${\bf 28}$.  Together, these charges transform in
the ${\bf 56}$ representation of $E_{7(7)}$ ; we can write combined charge vectors with eight of  these
charges turned on:
\begin{eqnarray}
\Gamma_p &=& \Big( Q_0^p,\; Q_i^p,\; Q^0_p,\; Q^i_p \Big) =
 \Big( {-L\over 2} n_p,\;\; {-1\over 4L} s_{ijk}
{d_p^jd_p^k\over n_p},\;\; {1\over 2L^2} {d^1_pd^2_pd^3_p\over
n_p^2}, \;\;
 {1\over 2} d^i_p\Big).
 \Label{chargevec} \\
 \Gamma &=& \sum_p \Gamma_p =
  \Big( Q_0,\; Q_i,\; Q^0,\; Q^i \Big)
  \Label{totalchargevec}
\end{eqnarray}
(To avoid clutter we leave out the D0, D2 etc. notation in
(\ref{quantcharge}).)
The symplectic $E_7$ invariant constructed from these
charges is
\be
<\Gamma_p,\Gamma_q> \, =\, {1\over 2}
\(Q^0_p Q_0^q - Q^0_q Q_0^p + Q^i_p Q_i^q - Q^i_q Q_i^p\)
 = {\Gamma_{pq}\over 8 L} = G^{(4)}_N\gamma_{pq},\; (\gamma_{pq} \in \Z).
 \Label{integerint}
\ee
Similarly, we can think of our eight harmonic functions as
part of a single harmonic function valued in the ${\bf 56}$ of $E_{7(7)}$
written as ($\Gamma_\infty$ denotes the constant terms):
\be
{\cal H} = \Big(
M_0,\; M_i, \; K^0, \; K^i \Big) = \Gamma_\infty + \sum_p {\Gamma_p\over
\rho_p} \, . \Label{harmvec}
\ee
Then the 1-form in (\ref{answer}) which gives rise to the
angular momentum in the solution satisfies
\be
\star_3 d(k_adx^a) = \star_3 d\vec{k} = \,<d{\cal H},{\cal H}>
\Label{angmom}.
\ee
At this juncture, we see that the reduction of our M-theory solutions down to
${\cal N}=8$ supergravity in four dimensions yields a framework similar to the
${\cal N}=2$ setup in \cite{denef1,denef2,denef3}. In fact, our theory with
eight vector fields and six scalars corresponds to a truncation of the ${\cal N}=8$
theory to the famous STU ${\cal N}=2$ model \cite{rhamfeld,DR1} which corresponds to the
symmetric coset space $[SL(2,R)/U(1)]^3$.

\paragraph{Explicit solution and special geometry: }
The four-dimensional metric is the obvious truncation of
(\ref{10dmetric}) \be ds^2_{4} = - J_4^{-1/2} (dt +k_a \, dx^a)^2
+ J_4^{1/2}  ds^2_{\R^3} \Label{4dmet} \ee
The four dimensional dilaton is constant, because the volume of the three
2-tori in (\ref{10dmetric}) cancels the spatial dependence of the ten
dimensional dilaton in (\ref{dilaton}). The solution has three complex scalar
fields  coming from the complexified Kahler moduli of the three $T_i$ ($\phi^i
= i{\rm Vol}^i + B_2^i$).  In terms of the explicit ten dimensional solution
(\ref{10dmetric},\ref{dilaton}), the scalars are
 \beq \phi^i &=&
i{-2J_4^{1/2}\over 2M_0M_i - s_{ijk}\,K^jK^k}- \({2M_iK^i + M_0K^0 - \sum_{j}
M_jK^j\over
2M_0M_i - s_{ijk}\,K^jK^k}\),\\
&=& \({\partial J_4^{1\over 2} \over \partial M_i} - {i\over 2}K^i\)\Bigg/
\({\partial J_4^{1\over 2} \over \partial K^0} + {i\over 2}M_0\) \, .
\Label{simplescalars}
\eeq
 In the first line there is no sum on $i$.  In the second line the scalars have
been written entirely in terms of the $J_4$ invariant and individual harmonic
functions transforming in an eight-dimensional subspace of the ${\bf 56}$ of
$E_{7(7)}$. These six scalars are part of the ${\bf 70}$ scalars of the
$E_{7(7)}/SU(8)$ coset. Finally, the $C_{3}$ field and $C_{1}$ field give $3+1$
vectors (the above three vector multiplets plus the graviphoton) which
transform, along with their duals, in the same subspace of the ${\bf56}$
representation as the harmonic functions. The potentials and their duals can be
summarized in a single vector as
\be {\bf{\cal \vec{A}}} = \Big(\vec{{\cal A}}_0,\;\vec{{\cal A}}_i,\;
\vec{{\cal A}}^0,\;\vec{{\cal A}}^i \Big),
\ee
with components from reducing
$C'_3$ and $C_1$:
\be \vec{{\cal A}}^i = - {1\over J_4}{\partial J_4\over
\partial M_i}(dt + k_a dx^a)
 + 2\vec{a}^i, \qquad
\vec{{\cal A}}_0 = + {1\over J_4}{\partial J_4\over \partial
K^0}(dt + k_a dx^a)
 + 2\vec{a}_0,
\ee
and their duals:
\be \vec{{\cal A}}_i = + {1\over
J_4}{\partial J_4\over \partial K^i}(dt + k_a dx^a)
 + 2\vec{a}_i, \qquad
\vec{{\cal A}}^0 = - {1\over J_4}{\partial J_4\over \partial
M_0}(dt + k_a dx^a)
 + 2\vec{a}^0.
\ee
The magnetic parts of these potentials are concisely written
as:
\be d\vec{a} = \Big( d\vec{a}_0,\; d\vec{a}_i,\; d\vec{a}^0,\;
d\vec{a}^i \Big)
  = - \star_3 d{\cal H}.
\ee

\paragraph{$E_{7(7)}$ and more general solutions: }
The  solutions presented above have four vector fields and six scalars,
corresponding to a truncation of the the full ${\cal N} =8$ theory to the
${\cal N} =2$ STU model \cite{DR1}. More generic four dimensional solutions with arbitrary
charges can be obtained by generalizing the above, allowing the harmonic
functions (\ref{harmvec}) as well as the indivdual charge vectors
(\ref{chargevec}) to occupy the full ${\bf 56}$ of $E_{7(7)}$. Typically,
however, the most general configurations of this type will break supersymmetry.
Since we are interested in 1/8-BPS states, our charges always need to line up
with a preferred ${\cal N} =2$ subalgebra, corresponding to the reduction of
$E_{7(7)}$ to $SO^{*}(12)\times SU(2)_R$ (for more details see
\cite{Andrianopoli:1997wi,Ferrara:2006em,FGK}).  For such configurations, the
appropriate charge subspace for supersymmetry yields vectors (and harmonic
functions) which transform in the real ${\bf 32}$ spinor representation of
$SO^*(12)$, accompanied by thirty non-trivial scalars.  The remaining forty
scalars of the ${\cal N}=8$ theory appear as hypermultiplet scalars with
respect to the $N=2$ truncation, parameterizing the coset
$E_{6(2)}/SU(2)_R\times SU(6)$.  Their values stay fixed throughout our
solutions, and are generally constrained by the alignment of the subgroup
$SO^{*}(12)\times SU(2)_R$ inside $E_{7(7)}$.

\paragraph{Generalized constraint and coarse graining: }
For general charge vectors $\Gamma_p$ the constraint equation (\ref{newbubble}) can be appropriately generalized by requiring integrability (\ref{angmom}) as before.  Then, in terms of a symplectic product of the charges and the asymptotic values of the harmonic functions (\ref{harmvec}),
\be
<\Gamma_p,\Gamma_\infty> + \sum_q {<\Gamma_p,\Gamma_q>\over \rho_{pq}} = 0.
\Label{newconstraint}
\ee
The first term automatically provides an expression for the $\Psi_p$'s and
summing these equations provides the first of the two constraints on
$\Gamma_\infty$,
\be
<\Gamma,\Gamma_\infty>=0,\qquad I_4\(\Gamma_\infty\) = 1,
\ee
which tell us that the asymptotic scalars sit in the appropriate coset.

In the form above, the constraint equation also gives us an insight into the
behavior of the solution if we ``coarse-grain'' over a collection of charges by
collecting them into a single pole.     For example, let us partition our poles
into sets ${\cal P}$ containing poles seperated by distances much smaller than
some reference scale, $\Lambda$.   Then we can define a coarse-grained solution
by assigning the total charge of each cluster ${\cal P}$ to the mean location
of the poles in the cluster:
\be
\Gamma_{\cal P} = \sum_{p\in {\cal P}} \Gamma_p,\qquad
\vec{x}_{\cal P} \sim < \vec{x}_p>.
\ee
The constraints on the coarsened solution are then:
\be
<\Gamma_{\cal P},\Gamma_\infty> + \sum_{\cal Q} {<\Gamma_{\cal P},\Gamma_{\cal Q}>
\over \rho_{{\cal PQ}}} \; <<\;  \Lambda.
\ee
An important feature of this coarse-graining is that if a cluster of poles
${\cal P}$ has the property that $<\Gamma_{\cal P},\Gamma_\infty> = 0$, then
this cluster can be placed at an infinite distance from the rest of the poles,
distinguishing our supergravity solution as one generated by at least two
separate bound states.

The maximally coarse-grained solution replaces the detailed microstate by a
solution with a single pole carrying the total charge vector of the spacetime.
Below we will show that such single-pole solutions are black holes of the $\CN
= 8$ supergravity and will in general have a finite horizon area.

\subsection{Relation to finite area black holes} \label{sec-finite}
We would like to understand how the solutions described above can be seen as
candidate microstates for the extremal black holes of  $\CN = 8$ supergravity.
Far away from all the poles, the harmonic function ${\cal H}$ can be well
approximated by the single center function
\be
{\bar{\cal H}}  = \Gamma_\infty + {\Gamma\over\rho}
\ee
where $\Gamma$ is the total charge vector defined in (\ref{totalchargevec}).
Plugging this simplified function into our formulism yields a metric with no
angular momentum~\footnote{Non-zero angular momentum is contingent on keeping
at least a dipole moment when approximating $\cal H$.} and
\be
J_4(\rho) = {I_4(\Gamma) \rho^{-4}}\,,
\ee
where $I_4(\Gamma)$ comes from expression (\ref{quartic1}) for $J_4$ after we
substitute in the appropriate charge for each harmonic function.
If $I_4(\Gamma) > 0 $, this is simply the spherically symmetric metric of a 1/8-BPS black hole
with total charge vector
$\Gamma$ and horizon area given by
\begin{equation}
A  = 2\pi \sqrt{I_4(\Gamma)} \, .
\end{equation}
If $I_4(\Gamma) < 0$ (the null case will become clear later), we have taken our coarse-graining
procedure too far, and need to retreat back until we only have centers with $I_4(\Gamma_{\cal P}) \ge 0$.
The astute reader can recognize $I_4$ as Cartan's quartic $E_7$ invariant
\cite{kalloshkol,vijaycargese,bw2,per2}. In
terms of the antisymmetric central charge matrix $z_{ij} = x_{ij} + i y_{ij}$
of $\CN = 8$ supergravity, this invariant is written as
\begin{equation}
4I_4 = {x^{ij}y_{ij} x^{kl} y_{kl} \over 4} - x^{ij}y_{kj} x^{kl} y_{li} - {1
\over 96} (\epsilon^{ijklmnop} x_{ij} x_{kl} x_{mn} x_{op} +
\epsilon_{ijklmnop} y^{ij} y^{kl} y^{mn} y^{op} ) \, .
\end{equation}
Here indices are raised and lowered by $\delta^i_j$.   In our
conventions\footnote{The $J_4$ invariant in \cite{vijaycargese}
differs by a factor of 4 from the one here, and charge conventions
there are related to the present ones by a factor of $\sqrt{2}$
and sign flip of the D6-brane charge.}
\begin{eqnarray}
x^{12} &=&  Q^1 ~~~;~~~ x^{34} =Q^2 ~~~;~~~ x^{56} =Q^3 ~~~;~~~ x^{78} = Q^0    \Label{xdef} \\
y_{12} &=& Q_1 ~~~;~~~ y_{34} = Q_2 ~~~;~~~ y_{56} = Q_3 ~~~;~~~
y_{78} = Q_0    \Label{ydef}
\end{eqnarray}
For a solution with only the D2 (D4) and D6 (D0) brane charges, the expected
horizon area of the associated black hole is
\begin{eqnarray}
{\rm D2-D2-D2-D6 \ area: } ~~~~~ A &=& 2\pi \sqrt{-Q_0 \prod_i
Q_i}
\Label{2226area} \\
{\rm D4-D4-D4-D0 \ area: } ~~~~~ A &=& 2\pi \sqrt{-Q^0 \prod_i
Q^i} \Label{4440area}
\end{eqnarray}
The sign under the square root indicates that the D6 (D0) must be appropriately
oriented relative to the D2 (D4) branes to preserve supersymmetry
\cite{ferraramald}.    If electric and magnetic dual objects are present, other
terms in $I_4$ will also contribute.  The space of all extremal finite area
black holes in $\CN = 8$ supergravity can be generated by $E_{7(7)}$
transformations of generating configurations with five charges containing one
electric-magnetic pair \cite{Andrianopoli:1997wi, ferraramald, vijaycargese}.
An example generating configuration contains three D2-brane, D6-brane and
D0-brane charges.    Such charge vectors are included in our analysis and thus
the solutions described earlier in this section provide candidate microstates
for the generating black holes of $\CN = 8$ supergravity.

It is interesting to ask whether a charge vector of the form
(\ref{chargevec}) associated to a single pole in the solution could have given rise to a finite horizon area
by itself.  Knowing the leading behavior of the functions $Z_i,k_0$ and $H$
it easy to show that
\begin{equation}
\lim_{\vec{x} \to \vec{x}_p} J_4 \propto \rho_p^{-1} \qquad
\Rightarrow\qquad I_4(\Gamma_p) = 0.
\end{equation}
We can also check that the invariant $I_4(\Gamma_p)$ vanishes in this case by
using the conventions (\ref{xdef},\ref{ydef}) and the
single pole charge vector as given in (\ref{chargevec}).
Thus, a black hole carrying the charges of a single pole in our
solutions (\ref{chargevec}) would have vanishing horizon area and
entropy.  In fact, the growth of $J_4$ near one of our
``primitive'' poles tells us even more.  In (\cite{ferraramald})
the authors distinguish four kinds of BPS black holes in ${\cal N}=8$
supergravity.  One can find examples for each of these by once again looking at
configurations with just D2 and D6-brane charge:
\beq
D2-D2-D2-D6\ne 0:\; 1/8 \,{\rm BPS},\, I_4 > 0,&& J_4 \propto \rho^{-4},
 \label{fourclasses} \\
D2-D2-D6\ne 0:\; 1/8 \,{\rm BPS},\, I_4 = 0,&& J_4 \propto \rho^{-3},
\nonumber \\
D2-D6\ne 0:\; 1/4 \,{\rm BPS},\, I_4 = 0\, , \partial I_4=0,
&& J_4 \propto \rho^{-2},
\nonumber \\
D6\ne 0:\; 1/2 \,{\rm BPS},\, I_4 = 0,\, \partial I_4=0, \,
\partial\partial |_{Adj}I_4 =0 ,&& J_4 \propto \rho^{-1} \nonumber .
\eeq
The notation with $\partial$'s is impressionistic, see \cite{ferraramald} for
more detail.  Thus, the rate of growth of $J_4$  yields a simple U-duality invariant method for determining how much supersymmetry is preserved by a given black hole or pole in a multi-pole configuration.

It has been shown on general grounds that to be associated to a finite horizon
area, the charge vector of a black hole in $\CN = 8$ supergravity can preserve
at most four supercharges -- such states, including solutions with general
2-brane and 6-brane charges are 1/8-BPS.  By contrast, the
classification above tells us that charge vectors of individual poles in our
solutions  are all 1/2-BPS.  Another simple way of seeing this is to note that
the charges in (\ref{chargevec}) are consistent with  having $-n_p$ 6-branes
with worldvolume fluxes turned on in the 12, 34, and 56 directions.   The
Chern-Simons couplings on the brane would then precisely reproduce the 4-brane,
2-brane and 0-brane charges given in (\ref{chargevec}).   Such states of
6-branes with fluxes are known to be 1/2-BPS, they are T-dual to IIB 3-branes
at angles on a $T^6$.

Previous work has shown how certain supersymmetric black holes could
be thought of as single center marginal bound states of 1/2 BPS
objects later understood to be D-branes \cite{DR1,DR2,DR3,DR4}.
Indeed the four classes of solutions in (\ref{fourclasses}) correspond
precisely to the four types of axion charges appearing in
\cite{DR1,DR2,DR3,DR4}. In our work, the component 1/2 BPS states are
spatially separated and are held together in a true bound state
because of their nonzero intersection numbers.

\paragraph{$E_{7(7)}$ transformations of microstates:}
In general, we can use an $E_{7(7)}$ transformation to take any finite area 1/8
BPS black hole of $\CN =8$ SUGRA with all ${\bf 56}$ charges to one with only
the eight charges of the STU model~\cite{Andrianopoli:1997wi,DR1}, with scalars in
$[SL(2,R)/SO(2)]^3$. As we mentioned above, we can further use the three
compact $SO(2)$'s of the STU models to eliminate\footnote{Actually, the full
quantum theory has a more restricted U-duality group which only allows us to
reduce the D4-brane number charges to be in the interval $0...(N_6-1)$, this is
the IIA manifestation of the fact that M-theory on Taub-NUT has $Z_{N_6}$ {\em
torsion} charges for M5-branes. } three more of the charges to get, for
example, a model with just D2-D2-D2-D6-D0 charges.    Recall, however, that the
proposed microstates for such a black hole will contain a large number of poles
with individual charge vectors.  Even if the overall black hole charges fall
within the STU sector, the charges of the component poles in the underlying
microstates will not be restricted in this way since there is not sufficient
symmetry to rotate each pole individually; they will typically all lie in
different subspaces of the spinor of $SO^*(12)$\ \cite{FGK}.

\subsection{Summary and Proposal}

In this section, we reduced a class of smooth candidate microstate
solutions for supersymmetric five dimensional black holes and rings to
microstates for four dimensional black holes with eight
charges.   These spacetimes were written as multi-center solutions in which each center served as a 1/2-BPS ``atom'' for building up the full configuration.   The bound state nature of the overall solution was maintained by the mutual non-locality of the charges which led to constraints on their relative positions.
This motivates a conjecture:
\begin{center}
\begin{minipage}[c]{0.85\textwidth}
{\it
Every supersymmetric 4D black hole of finite area,
preserving 4 supercharges, can be split up into microstates made of primitive
1/2-BPS ``atoms", each of which preserves 16 supercharges.   In order to describe a bound
state, these atoms should consist of mutually non-local charges.
}
\end{minipage}
\end{center}
The next section provides evidence for this picture.

\section{From spacetime foam to D-branes}

In \cite{juanandy4d,JohnsonKhuriMyers,VijayFinn1,klebtseyt,MSW} the entropy of
four dimensional black holes of finite area was accounted for in terms of
D-brane  bound state degeneracies.   The basic strategy was to use D-branes to
count the states in the $g_s \to 0$ limit and then extrapolate back to stronger
coupling using supersymmetry.   As such the microstructure of the black hole
arose from the many degenerate ground states of the D-branes wrapped on the
internal space, in our case $T^6$.    The picture offered above suggests
instead that the entropy of the black hole arises from structure in the
non-compact four dimensions of spacetime.   This is more along the lines of the
proposals of Mathur and collaborators
\cite{mathuradsparadox,mathurstretch,mathurproposalessay}.  In this section we
demonstrate how these two approaches to black hole entropy could be reconciled.
The classical supergravity solutions that we have found flow at weak coupling
to systems of intersecting branes of the sort originally used in
\cite{juanandy4d,JohnsonKhuriMyers,VijayFinn1,klebtseyt,MSW}
 to count black hole microstates.
This suggests  that the ground states of such coincident branes at vanishing
$g_s$ flow at weak $g_s$  to the four dimensional supergravity solutions that
we have described and at strong $g_s$ to a spacetime foam in M-theory.
Evidence for this picture exists in the analysis of flows of BPS brane bound
states discussed by Denef \cite{denefhall}.

\subsection{A scaling relation}
\Label{scalingsection}

The locations of the centers in each four dimensional geometry in
Sec.~3 are determined by the constraint equation (\ref{newbubble}).
We can examine how these solutions change as
\begin{equation}
g_s \to \beta g_s
\Label{gsscale}
\end{equation}
while we hold the volume of the torus fixed in string units
\begin{equation}
{\prod_i V_i \over l_s^6} = {\rm fixed} \, .
\end{equation}
The quantized charges of the branes, constructed from the integers
$n_p$ and $m_p^i$ (\ref{integerquant}) are held fixed, and thus the
physical charges (\ref{quantcharge}) scale in powers of $\beta$.
Putting everything together, under the rescaling (\ref{gsscale}), the
constraint equation (\ref{newbubble}) becomes
\begin{equation}
4 \Psi_p \beta + \beta^3 \, \sum_q {\Gamma_{pq} \over r_{pq}} = 0 \, .
\end{equation}
Thus, given a set of separations $\{r_{pq}\}$ that solve the
constraint equation for a string coupling $g_s$, the set of
separations $\{ \beta^2 r_{pq} \}$ solves the constraints for a
coupling $\beta g_s$.  The physical separations are
\begin{equation}
\rho_{pq} = {2 r_{pq} \over L} \Label{rhoscaling}
\end{equation}
and these scale linearly with $g_s$:
\begin{equation}
\rho_{pq} \to \beta \rho_{pq} .
\end{equation}
In order to continue to satisfy the no-CTC condition (\ref{noCTC}) we must also scale the coordinate $\rho$ as
\begin{equation}
\rho \to \beta \rho .
\end{equation}
This scaling has far reaching consequences. As we go to weaker
coupling, the branes move closer together in string units. At some
value of the coupling the branes will move within a string length of
each other and the appropriate description is in terms of the open
strings on the branes.

To estimate when the open string picture becomes valid, first consider
the two-center case.  Define the dimensionless quantity \beq \psi_p =
(g_s l_s)^{-1} \Psi_p \ .  \eeq The branes will be much closer than a
string length apart when\footnote{Recall in the two-pole case, with
$\alpha=0$, the constraint (\ref{extra}) sets
$\Psi_1=-\Psi_2$.}  \beq g_s \ll {1 \over 32 \pi^6} { \prod_i V_i
\over l_s^6} \left| {\psi_1 \over \gamma_{12} } \right| .  \eeq When
all brane separations are roughly of the same order we can estimate
that all pairs of branes are closer than the string length when \beq
g_s \ll {1 \over 32 \pi^6} { \prod_i V_i \over l_s^6} \left| {\psi_p
\over \gamma_{pq} } \right| ~~~~ \forall \, \psi_p,\gamma_{pq} \, .
\eeq For a general bound state, there will always be a value of $g_s$
small enough that all branes are separated by distances smaller than
the string length.  For such small $g_s$ the supergravity solutions
described in Sec. 3 are better described in terms of open string
degrees of freedom coming from the D-branes.

Going the other way, as $g_s$ increases, the intervals between the IIA
D-branes in the solution increase until, at large coupling the IIA
description is no longer valid.  At that point we move to an M-theory
description with the multi-center D-brane solution replaced by a
network of two-cycles (``bubbles") we call spacetime foam. The new
reference length becomes the 11-dimensional Planck length; the size of
the asymptotic circle in Planck units now becomes a dimensionless
modulus like all the others.  Recently, a similar flow from D-branes
to ``spacetime foam'' has been noted in the topological string
\cite{DVVtop}.

\subsection{The open string picture}
\Label{sec-open}

When D-branes are much closer than a string length, an open string
description is appropriate.  The vacua of the brane system can be
analyzed just in terms of the low-energy field theory on the D-branes
if the massive string excitations can integrated out.  This is the
case when all the brane separations are less than $l_s$ and when all
the charge vectors are sufficiently closely aligned.  Taking these
conditions to be satisfied, we will describe how the supergravity
states in Sec. 3 appear as vacua of a D-brane gauge theory when $g_s$
is sufficiently small.  In this section, we will set $l_s =1$.

First consider a single center with charge vector $\Gamma_p = N_p \,
\hat{\Gamma}_p$ where $N_p$ is the greatest common divisor of the
charges appearing in $\Gamma_p$, and $\hat{\Gamma}_p$ is thus a {\it
primitive} charge vector.  This represents a stack of $N_p$ D-branes
wrapping $T^6$.  When the torus is small (as we take it to be), the
low energy physics is obtained by dimensionally reducing the D-brane
gauged field theory to a gauged quantum mechanics.  Thus the latter
has the field content of dimensionally reduced $\CN = 4, d=4$,
$U(N_p)$ super-Yang-Mills theory.  However, since the interactions
between different stacks of branes will only preserve four
supercharges, it is convenient to decompose into multiplets of $\CN=4,
d=1$ Yang-Mills, obtained via dimensional reduction of an $\CN =1,
d=4$ theory.  This gives one vector multiplet and three adjoint chiral
multiplets.  The vector multiplet contains three real, adjoint scalars
-- these parametrize the positions of the stack of branes in the
non-compact space.  The three complex adjoints parametrize Wilson
lines on the 6-branes (or positions within the torus after
T-duality).\footnote{Under T-duality of the torus, the 6-branes we are
describing can be transformed to a system of wrapped 3-branes.  The
worldvolume theory of these branes consists of a $\CN=4, d=4$ vector
multiplet which decomposes into a vector multiplet of $\CN=1, d=4$
Yang-Mills and three adjoint chiral multiplets.  The real parts of
these adjoint scalars parameterize positions in the non-compact space,
while the imaginary parts describe positions on the torus.  Upon
dimensional reduction, the real parts of the $d=4$ adjoints become the
three real scalars in the $\CN=4,d=1$ vector multiplet.  The imaginary
parts of the $d=4$ adjoints pair up with the Wilson lines on the
3-brane to build the three adjoint chiral multiplets of the $d=1$
theory.}

In addition, stacks of branes with integer intersection number $\gamma_{pq}
\neq 0$ (\ref{integerint}) give rise to  $|\gamma_{pq}|$ chiral fields
$\phi_{pq}$ charged in the bifundamental of $U(N_p) \times U(N_q)$
\cite{quivers,branesatangles1,branesatangles2}.     If $\gamma_{pq} = 0$, i.e.
if the charge vectors $\Gamma_p$ and $\Gamma_q$ are mutually local, then
bifundamentals for these branes only appear in chiral and anti-chiral pairs.
There is a Higgs term coupling these bifundamentals to the massless adjoint
multiplets on each of the two branes. In a generic bound state, however,
mutually local charge vector pairs will occur rarely. Hence we will focus on
situations where $\gamma_{pq} \neq 0 \, \forall p, q$.

Thus, the low-energy field theory describing the system of branes at small $g_s$ is a $d=1, \ {\cal N}=4$ quiver quantum mechanics constructed as follows:
\begin{enumerate}
\item{For each stack of D6-branes at point $p$ with charge vector $\Gamma_p$ we associate a $U(N_p)$ gauge group and vector multiplet.  Here $\Gamma_p = N_p \, \hat{\Gamma}_p$ where $N_p$ is the GCD of the charges in $\Gamma_p$ and $\hat{\Gamma}_p$ is {\it primitive}.
The three real, adjoint scalars ($\vec{x}$) in the vector multiplet  correspond to location in the $\R^3$ transverse to the branes.}
\item{The number of bifundamental chiral fields $\phi_{pq}$ (transforming in the $(\bar{N_p},N_q)$ of $U(N_p) \times U(N_q)$)
between branes $p$ and $q$ is given by the intersection number $\gamma_{pq}$.}
\item{Since each charge vector $\Gamma_p$ corresponds to a 1/2-BPS state, we will
also get at each node the remainder of an ${\cal N}=4$ vector multiplet:
three adjoint chiral multiplets.  These will mostly play a spectator role in our considerations.}
\end{enumerate}

\begin{figure}
\centering \hspace{0.2in} \includegraphics[width=0.4\textwidth]{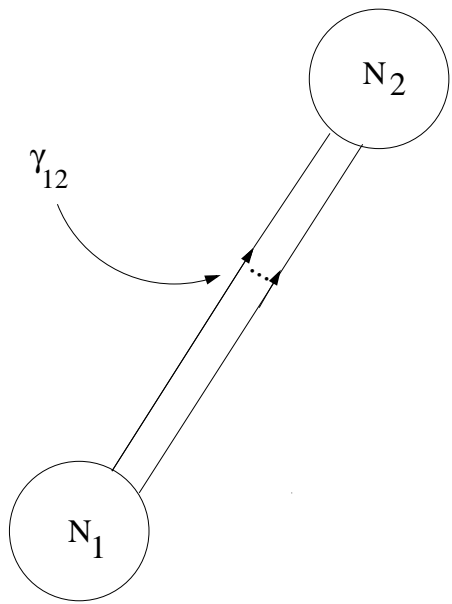} \caption{A sample quiver with two nodes.}
\end{figure}

The general Lagrangian for the vector multiplet, the bifundamental chiral
multiplet and their couplings is given in Appendix C of \cite{denefhall}. The
terms arising from the additional chiral adjoints can be determined by
dimensional reduction of the $\CN=4, d=4$ Lagrangian.  For determining the
vacua and phases of the theory, we need to know how these additional fields
contribute to the D-term and F-term equations and to the masses of the
bifundamentals.   The FI-term in the
Lagrangian  takes the form \cite{denefhall}
\begin{equation}
L_{FI} = \sum_p \CF_p \, {\rm Tr}D_p
\end{equation}
where $D_p$ is the auxiliary adjoint field in the  $U(N_p)$ vector multiplet.
$\CF_p$ is linear in the charges and depends on  closed string field
values.  It only couples to the D-term for the diagonal $U(1)$ of $U(N_p)$.
This is consistent with the notion that if we just slightly separate our
stack into two stacks, $U(N_p)\to U(N_{p_1})\times U(N_{p_2})$, then
$\CF_p=\CF_{p_1}=\CF_{p_2}$.

Since the adjoint chirals are neutral under the diagonal center-of-mass $U(1)$
they cannot couple to the corresponding D-term and hence to the D-term equation
which will most interest us; they only contribute to the other $N_p -1$
equations coming from the $SU(N_p)$ D-terms.  What is more, up to at least
quadratic order, the adjoint scalars do not have Higgs couplings to chiral
bifundamentals which are not paired to anti-chiral ones.  To see this, consider
a pair of branes with bifundamentals running between them. The pair can be
T-dualized to give two 3-branes at angles in IIB string theory. In this
context, the expectation values of the adjoint scalars in the quantum mechanics
Lagrangian parameterize the positions of the 3-branes on $T^6$ and the Wilson
lines in the branes. Neither of these affect either the number of intersections
between the branes, or the spectrum of strings localized at the intersection
points.  Hence there are no Higgs couplings up to at least quadratic order
between the adjoint and the bifundamental chiral fields. Finally, the
potential for the adjoint chiral multiplets is inherited from $\CN=4, d=4$
Yang-Mills and simply forces the adjoints to commute on the vacuum manifold.

Hence, for the purpose of studying the phases and vacua of our quiver quantum
mechanics, we can largely ignore the adjoint chiral multiplets.   Fortunately, the
remaining problem is identical to the one studied by Denef in \cite{denefhall}
and in Sec.~\ref{gaugeanalsec} we simply adapt his analysis to our situation.
The vacuum manifold of the quiver quantum mechanics can have a Coulomb branch
in which the vector multiplet scalars ($\vec{x}_p$) have expectation values,
and a Higgs branch in which the chiral multiplet scalars $\phi_{pq}$ are given
VEVs.   We will study each in turn and discuss how the supergravity states in
Sec. 3 appear in the Coulomb branch and how they can flow into the Higgs branch
as $g_s \to 0$.

\paragraph{An example black hole microstate: }   Before proceeding it
is worthwhile to give an example showing that quivers exist with
charges appropriate for being microstates of black holes with finite
area.  Since each D-brane center is 1/2-BPS we will require a minimum
of three nodes in the quiver (see Sec.~\ref{sec-finite}). The
conditions to be satisfied are:
\begin{enumerate}
\item All charges and charge vectors must be appropriately quantized
(1/2-BPS in the case of individual centers).
\item $J_4(\Gamma) > 0$ where $\Gamma$ is the total charge vector so
that the collection has the charges to be a candidate microstate for a
finite horizon area black hole.
\item There exist solutions of the constraint equations
(\ref{newconstraint}) that also satisfy the triangle inequalities for
brane positions.  This is the only one of our conditions which depends on the asymptotic
moduli.

\end{enumerate}
Working in units such that $e^i =1 , \, \forall i$ and $L=1$ (this
sets $G_N^{(4)} = 1/8$), the charge vectors can be written $\Gamma_p =
{1\over2}(N_0^p, N_i^p, N^0_p, N^i_p), \, i = 1 \ldots 3$
(\ref{chargevec}), where as before the $N$ charges are quantized.  We
will also specialize to diagonal 2 and 4-brane charges
(i.e. $N^i_p=N_p, \ N_i^p=N^p, \quad \forall i$). An example quiver
meeting our requirements arises from the charge vectors
\begin{equation}
\Gamma_1 = {1\over2}(6,0,0,0) , \ \ \Gamma_2 = {1\over2}(-1,-1,1,1), \ \ \Gamma_3={1\over2}(-2,-2,-2,-2) .
\end{equation}
The total charge vector and $J_4$ invariant are
\begin{equation}
\Gamma= \sum_p \Gamma_p = {1\over2}(3,-3,-1,-1) ~~~~~;~~~~~  J_4(\Gamma) = 71/16 = 71\cdot (2G^{(4)}_N)^2>0 \, .
\end{equation}
The intersection numbers
\begin{equation}
\gamma_{12}=-6, \ \ \gamma_{23}=-16, \ \ \gamma_{31}=-12,
\end{equation}
indicate that we have closed loops in the quiver. 

For such a closed loop there exists a simple ``scaling solution''\cite{denefhall} where the centers
converge on each other with separations limiting to a set congruent with the triangle made up of 
the $\gamma_{pq}$'s.  This is independent of the value of the asymptotic moduli set by our choice
of $\Gamma_\infty$.

\subsection{Gauge theory analysis}
\Label{gaugeanalsec}

The results of \cite{denefhall} are expressed in the language of $\CN = 2$
supergravity,in terms of the central charge associated to each brane in the
quiver.   In Appendix A the central charge of the p$^{{\rm th}}$ brane is shown
to be
\begin{equation}
Z_p = {L \over 2} n_p - {1 \over 2L} \sum_{i,j,k} {s_{ijk} \over 2}
n_p \lambda^j_p \lambda^k_p +{i \over 2} \Psi_p . \Label{centralcharge}
\end{equation}
In terms of $Z_p$ the mass of the brane is
\begin{equation}
m_p = {|Z_p| \over 4 G_4}
\end{equation}
and we can write
\begin{equation}
Z_p = |Z_p| e^{i\alpha_p} ~~~~;~~~~ \sin\alpha_p = {{\rm Im} Z_p \over |Z_p|}
\end{equation}
The total central charge is
\begin{equation}
Z = \sum_p Z_p ~~~~~;~~~~~ Z \equiv |Z| e^{i \alpha} ~~~~~;~~~~~
\sin\alpha =  {{\rm Im} Z \over |Z|}
\end{equation}
In our solutions the constraint (\ref{extra}) leads to $\alpha  =
0$.\footnote{A  non-zero overall phase is easily restored by including
solutions that carry a velocity along the Taub-Nut direction, i.e. by allowing
a total charge vector $\Gamma$ such that $\tilde\alpha \ne 0$.} In the field
theory analysis below the parameter
\begin{equation}
\theta_p = {{\rm Im}(e^{-i\alpha} Z_p) \over 4 G_4} = |Z_p| {\sin(\alpha_p - \alpha) \over 4 G_4} \approx |Z_p| {(\alpha_p - \alpha) \over 4 G_4}
\end{equation}
will play a role.   The last approximate equality holds when the phases of all the branes are nearly equal, as required for a field theory analysis to be valid.   Since we are working with solutions with $\alpha = 0$, this means that all the $\alpha_p$ are small also.  Thus
\begin{equation}
\sin\alpha_p \approx \alpha_p = {{\rm Im} Z_p \over |Z_p|}
\end{equation}
Putting everything together,
\begin{equation}
\theta_p \approx {\Psi_p \over 8 G_4} \, .
\Label{thetaresult}
\end{equation}
Armed with these quantities we can adapt Denef's results \cite{denefhall} to our setting.

We will consider quivers  that do not contain closed loops.  This ensures that the bifundamental chiral multiplets do not have a superpotential and hence analyzing the D-term equations is sufficient to determine the vacuum structure.  Also, for simplicity, we will  begin by taking $N_p = 1$ (a $U(1) $ gauge theory) at each quiver node.   The non-Abelian case will follow from this.   For an abelian quiver ($N_p = 1, \forall p$) without closed loops the relevant part of the bosonic effective Lagrangian is \cite{denefhall}:
\begin{equation}
L_{{\rm eff}} = \sum_p \left[ {m_p \over 2} D_p^2  - \theta_p D_p \right]
+
\sum_{p<q} \sum_{a=1}^{|\gamma_{pq}|}
\left[ |F_{pq}^a|^2 - \left( |\vec{x}_p - \vec{x}_q|^2 +
(D_p - D_q) (-1)^{s_{pq}}\right)   |\phi_{pq}^a|^2\right] \, .
\Label{efflag}
\end{equation}
Here $\vec{x}_p$ and $D_p$ are the three scalar fields and the auxiliary field of the p$^{{\rm th}}$ vector multiplet, $\phi_{pq}^a$ are the $|\gamma_{pq}|$ bifundamentals charged under $U(1)_p \times U(1)_q$, $F^a_{pq}$ are the corresponding auxiliary fields, and
\begin{equation}
s_{pq} = {\rm sign}(\gamma_{pq}) \, .
\end{equation}
We have left out the standard kinetic terms and fermionic pieces.   If some of the $N_p > 1$,  additional commutator terms and appropriate traces are required.

\paragraph{Coulomb branch: }  When the vector multiplet scalars are given an expectation value, the bifundamental fields between the branes at $p$ and at $q$ have a mass
\begin{equation}
(m^\phi_{pq})^2 =   |\vec{x}_p - \vec{x}_q|^2 + (D_p - D_q) (-1)^{s_{pq}}
\end{equation}
The fermionic partner of $\phi^a_{pq}$ has a mass (see Appendix C of \cite{denefhall})
\begin{equation}
(m^\psi_{pq})^2 =  |\vec{x}_p - \vec{x}_q|^2
\end{equation}
Thus the fields in the chiral multiplet can be integrated out to give an effective Lagrangian for the fields in the vector multiplet.   We are particularly interested in terms that are linear in $D_p$ since these make up the Fayet-Iliopoulos parameter whose vanishing gives the condition for supersymmetry.  Fortunately, a non-renormalization theorem guarantees that this will be exact at one-loop.   The bosonic effective Lagrangian for the vector multiplet at one-loop order in the chiral multiplet is
\begin{equation}
L_{{\rm eff}}^V =
\sum_p \left[ {m_p \over 2} D_p^2  - \theta_p D_p \right]
+
\sum_{p<q} \sum_{a=1}^{|\gamma_{pq}|}
\ln \det
\left[
-\partial_t^2 + |\vec{x}_p - \vec{x}_q|^2
\over
-\partial_t^2 + |\vec{x}_p - \vec{x}_q|^2  +  (D_p - D_q) (-1)^{s_{pq}}
\right]
\end{equation}
The determinants are standard and give
\begin{equation}
L_{{\rm eff}}^V =
\sum_p \left[ {m_p \over 2} D_p^2  - \theta_p D_p \right]
+
\sum_{p<q} |\gamma_{pq}| \left(   |\vec{x}_p - \vec{x}_q|- \sqrt{ |\vec{x}_p - \vec{x}_q|^2 + (D_p - D_q) (-1)^{s_{pq}} }\right)
\end{equation}
The D-term equation $({\partial L_{{\rm eff}}^V  / \partial D_p})|_{D_p = 0} = 0$
gives
\begin{equation}
\sum_q {\gamma_{pq}
\over
2 \,  |\vec{x}_p - \vec{x}_q|} = - \theta_p \, ,
\Label{coulombDterm}
\end{equation}
which, when combined with  (\ref{thetaresult}),  gives the supersymmetry conditions
\begin{equation}
\sum_q {\gamma_{pq} \over |\vec{x}_p - \vec{x}_q|} = - {1 \over 4 G_4} \Psi_p  \, .
\Label{coulombvacua}
\end{equation}
The solutions to this equation form the moduli space of supersymmetric vacua in the Coulomb  branch.     Now recall that our supergravity solutions satisfy a constraint equation
$\sum_p {\Gamma_{pq} /r_{pq}} = - 4\Psi_p $
(\ref{newbubble}).
Recalling the relation (\ref{integerint}) between $\Gamma_{pq}$ and the integer intersection numbers $\gamma_{pq}$, as well as the relation (\ref{rhoscaling}) between $r_{pq}$ and the physical separations $\rho_{pq}$, the supergravity constraint becomes
\begin{equation}
\sum_p {\gamma_{pq} \over \rho_{pq}} = - {1\over 4 G_4} \Psi_p \, .
\Label{sugravacua}
\end{equation}
It is beautiful that (\ref{coulombvacua}) and (\ref{sugravacua}) match
identically. This precise match
teaches us that, following the scaling relation in Sec.~\ref{scalingsection},
as $g_s$ decreases each supergravity solution in Sec.~3 flows smoothly into a
corresponding solution in the gauge theory Coulomb branch.\footnote{Strictly speaking,  in
situations where some brane separations are much larger than others, parts of
the solution can flow into the Coulomb branch while other parts remain better
described in supergravity.}

\paragraph{Higgs branch: }  The scaling relation in Sec.~\ref{scalingsection} applies equally to the Coulomb branch equation (\ref{coulombvacua}).   Hence, after our states have flowed into the Coulomb branch, a reduction of $g_s$ will cause a further decrease in $|\vec{x}_p - \vec{x}_q|$, and with it the mass of the chiral multiplet.   If this mass becomes too small, the field cannot be integrated out.  To study when this happens, we can eliminate the auxiliary field $D_p$ from (\ref{efflag}) via its equation of motion and find the mass of $\phi_{pq}^a$:
\begin{eqnarray}
(m_{pq}^\phi)^2 &=& |\vec{x}_p - \vec{x}_q|^2 + \left({\theta_p \over m_p} - {\theta_q\over m_q}\right)\,  (-1)^{s_{pq}} \, \nonumber\\
&=& |\vec{x}_p - \vec{x}_q|^2 + 4 G_4 (\alpha_p - \alpha_q)\,  (-1)^{s_{pq}} \,
.
\end{eqnarray}
For charge vectors admitting bound states one can show from (\ref{newbubble}) that for every $p$ there is at least one $q$ such that $(\theta_p - \theta_q) (-1)^{s_{pq}} < 0$.\footnote{One can readily argue that if  for some $p$, $(\theta_p - \theta_q) (-1)^{s_{pq}}  \geq 0, \forall q$ then there is no solution to the constraint equation (\ref{newbubble}) or (\ref{coulombvacua}) for finite $r_{pq}$.}  For such pairs, the mass of the bifundamentals $\phi_{pq}^a$ will vanish and then become negative when $|\vec{x}_p - \vec{x}_q|$ is sufficiently small.   In view of the scaling relation in Sec.~\ref{scalingsection}, this means that as  $g_s \to 0$ some of the bifundamental chiral multiplets will become massless and then condense, taking the theory into the Higgs branch.   Near the condensation point these fields are light and cannot be integrated out as in the analysis of the Coulomb branch.  Indeed, since we are dealing with a one-dimensional effective Lagrangian, the wavefunction of a state can have a spread that overlaps the classical Higgs and Coulomb branches.\footnote{Such overlaps were discussed in various contexts in \cite{denefhall,michaherman,wittenphase,wittenconformal}.}   We will not attempt the full quantum mechanical treatment of the wavefunction here (see \cite{denefhall} for some details) and instead analyze the classical Higgs branch that arises when the $\phi^a_{pq}$ have large VEVs.

In the classical Higgs branch,  the vector multiplet scalars are set to zero -- they acquire a mass from the Higgs mechanism and can be integrated out.   From (\ref{efflag}), the D-term equation $({\partial L_{{\rm eff}}^V  / \partial D_p})|_{D_p = 0} = 0$ for supersymmetry  gives the condition
\begin{equation}
\sum_q \sum_{a=1}^{|\gamma_{pq}|} |\phi_{pq}^a|^2 (-1)^{s_{pq}} = - \theta_p, \ \ \forall p ,   \Label{HiggsDterm}
\end{equation}
The solutions to this equation define the Higgs branch vacuum manifold. For
example, if the quiver only has two nodes the vacuum manifold is
$CP^{|\gamma_{pq}| - 1}$.    In general we obtain some intersection of complex
projective spaces.    A simple ansatz for solving these equations is to take
all the bifundamentals between nodes $p$ and $q$ to have the same VEV
\begin{equation}
\phi^a_{pq} \equiv {1 \over 2\, R_{pq}} \, , \ \ \forall a \, . \Label{ansatz}
\end{equation}
Then (\ref{HiggsDterm}) becomes
\begin{equation}
\sum_q {\gamma_{pq} \over 2 \,R_{pq}} = -\theta_p \Label{CoulombInHiggs}
\end{equation}
Remarkably, this precisely reproduces the vaccum equation in the Coulomb branch (\ref{coulombDterm}) and the constraint equation in supergravity (\ref{sugravacua}), suggesting how the classical moduli space of solutions can flow between these phases as $g_s$ changes.

\paragraph{Matching the Coulomb and Higgs branches: }
At face value the classical Coulomb and Higgs branches each contain data that
is absent in the other.  In the Higgs branch, the $\phi^a_{pq}$ can each have
independent VEVs and the ansatz (\ref{ansatz}) seems to only explore a simple
subspace of the moduli space which reproduces the Coulomb branch.   In the
Coulomb branch the bifundamentals have been integrated out and the only piece that
remains from their data are the multiplicities $|\gamma_{pq}|$.   On the other
hand, in the Coulomb branch, any solution to the constraints (\ref{sugravacua})
must additionally satisfy triangle inequalities for the VEVs  $\rho_{pq} =
|\vec{x}_p - \vec{x}_q|$ ($\rho_{pq} + \rho_{q l} \geq \rho_{pl}$).   These
additional consistency conditions on a solution arise because the $\vec{x}_p$
transform as vectors of $SO(3)$, the four dimensional rotation group.    It is
important to understand how such triangle inequalities can arise in the Higgs
branch, since the D-term equations do not imply them.    The bosonic fields
$\phi^a_{pq}$ whose VEVs define the Higgs vacuum manifold are invariant under
$SO(3)$.   However, the fermionic partners of $\phi$, produced by the action of
a supercharge on $\phi^a_{pq}$, transform in a spinor of $SO(3)$.  This
suggests that there is a further consistency condition on the Higgs branch
vacua involving these fermions and the bosonic VEVs, but we have not identified
this condition here.    In addition,  there is an $SU(2)$ action, called the Lefschetz $SU(2)$, on the cohomology  of any K\"{a}hler moduli space.   The latter is determined completely by the defining equation of the variety (\ref{HiggsDterm}).    Relating the Lefschetz $SU(2)$ to the spatial rotation group in the Coulomb branch, it seems possible that the triangle inequalities appear as some kind of global integrability condition on the manifold specified by values of
$\phi^a_{pq}$ solving the D-term equations.

If the vacua in the Higgs and Coulomb branches can flow into each other as we
are advocating, a minimal requirement is that the {\it number} of vacuum states
in each branch should be equal.  In the Higgs branch we must count the ground
states of supersymmetric quantum mechanics on the classical moduli space
(\ref{HiggsDterm}).  These are well-known to be in correspondence with the
Dolbeault cohomology classes of the moduli space.  Thus the number of
supersymmetric ground states in the Higgs branch equals the sum of Betti
numbers of the moduli space.  For quivers without closed loops (and without
extra adjoint matter) there is a formula from Reineke that computes these
\cite{Reineke}. The corresponding problem in the Coulomb branch involves
quantizing the motion of charged particles in the presence of monopoles
(mutually nonlocal charges), and counting the resulting Landau levels.     This
has been done in some cases by Denef \cite{denefhall} and exactly matches the
count of states in the Higgs branch. Interestingly, identical particle-monopole
problems have appeared in recent approaches to counting the states of black
holes and in the relation of such counting problems to topological string
theory \cite{strom1,D4D0quintic,antid2}.

Finally, in our analysis we have separately studied the classical Higgs and Coulomb branches.  In order
to explicitly see a flow between them, we should construct the wavefunction in
our quantum mechanical system and observe how it flows with changes of the
coupling.   Again, we refer to \cite{denefhall} for a detailed analysis in
instructive special cases.

\paragraph{Non-abelian generalization and including adjoint chiral fields: }
We have focused on the case where all the $N_p=1$. For
more general values of $N_p$, we need to take a look at the effect of including
the non-abelian $SU(N_p)$ degrees of freedom.  For each node in our quiver, we
split the set of $N_p$ independent D-term equations into a singlet equation
corresponding to the D-term in the center-of-mass $U(1)$ and $N_p -1$ extra
equations coming from the remaining $SU(N_p-1)$ D-terms.  The singlet equation
is the only one which includes the FI term ($\CF_p$) and the adjoint scalars ($X^i_p$, $i=1,2,3$) do
not appear; solving the singlet equations will thus involve exactly the same exercise as
before.  The $SU(N_p)$ equations take the form,
written using the generators $t^\alpha_p$ ($\alpha = 1\ldots N_p^2 -1)$ \cite{wittenD0D6}:
\be%
\forall \alpha: \qquad \sum_q \sum_{a=1}^{|\gamma_{pq}|} Tr \left( t^\alpha_p \phi^{a}_{pq} \phi^{a\dag}_{pq} \right) (-1)^{s_{pq}}= - \sum_{i=1}^3 Tr \left(t^\alpha_p [X^i_p, \bar{X}^i_p]^2 \right) ,
\label{newDterms}
\ee
These additional equations reflect the fact that a
collection of $N_p$ of our ``atoms" is only classically situated at the
single center which our singlet D-term equations see; quantum mechanically the
$N_p$ identical branes form a cloud of particles whose features are controlled by the same matrix Lagrangian that describes the interactions of $N_p$ D0-branes \cite{DKPS,Matrixtheory}.
One of the key features of the $D0$-brane matrix Lagrangian is the condition that
\be
[X^i,X^j] =0 \qquad \forall i\ne j.
\ee
which comes from minimizing the potential for the adjoint scalar fields.
 Since the adjoint chiral fields couple to the chiral bifundamentals through (\ref{newDterms}) we expect the bifundamentals to affect
the internal dynamics of each ``cloud" of $N_p$ particles in some significant
fashion; perhaps the perturbations in each cloud will be correlated via the
singlet equations. This analysis is beyond the scope of our
discussion, but we expect that the $SU(N_p)$ degrees of freedom will be important in giving the black hole its finite entropy and spatial size.  This is an important difference in perspective compared
to~\cite{denefhall}.
Finally, note that when $N_p >
1$, it is easy to see, at least in the two center case fully discussed in~\cite{denefhall},
that the overlap between the Higgs and Coulomb phases increases.  This is
reminiscent of the ideas in~\cite{michaherman}.

\paragraph{Superpotentials: }
The main limitation of the analysis presented above is that it does not include
quivers with closed loops (e.g. Fig.~2).   Such quivers are generic amongst
black hole microstates, and will give rise to a superpotential for the chiral
multiplets.   While techniques for computing this superpotential for branes
wrapped on a torus are available \cite{Cremades:2003qj}, the computation is
done on a case by case basis, and will produce cubic and higher terms in the
chiral multiplets.   In the Coulomb branch these fields are massive, and the
effective action is computed by integrating them out.  Fortunately the
contribution to the D-term equation from this computation is exact at one-loop
and thus the superpotential plays no role in determining the Coulomb branch
moduli space. Thus our description of a flow between supergravity and a gauge
theory Coulomb branch as $g_s$ is varied is unchanged.   On the Higgs branch,
however, a superpotential $W$ will lead to a set of additional constraints,
namely $\partial W/\partial \phi_{pq}^a = 0$, within the manifold defined by
the D-term equation (\ref{HiggsDterm}).     While this will reduce the
dimension of the classical Higgs moduli space, the number of quantum states
could increase, decrease or remain unchanged depending on the cohmology of the
constrained manifold.     Unlike the case without closed loops \cite{Reineke},
a general formula for the number of states in the Higgs branch is not available
and hence the relation to the Coulomb branch in this case remains to be
studied.  In particular, while the analysis in \cite{denefhall} and above shows
that states in the Coulomb branch will flow into the Higgs branch at very weak
coupling, the converse is not obvious.

\begin{figure}
\centering \hspace{0.2in} \includegraphics[width=0.5\textwidth]{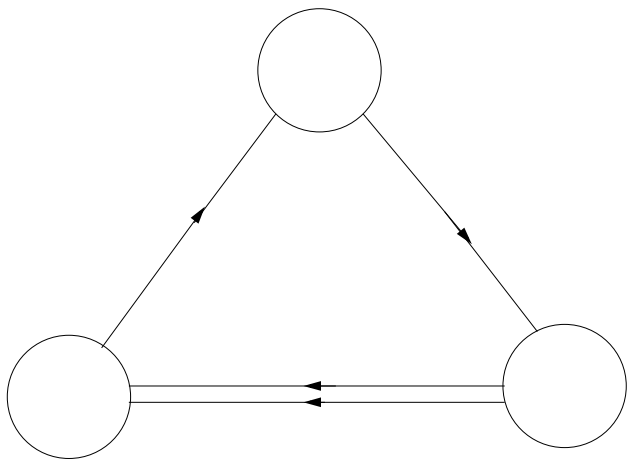} 
\caption{A quiver with a closed loop.}
\end{figure}

\subsection{Summary and proposal}

In \cite{denefhall}, Denef suggests that states that flow from the Higgs branch
into the Coulomb branch as the coupling is increased form a class of
multi-center solutions separate from the black hole solutions that describe a
large multiplicity of Higgs branch microstates.      His reasons for suggesting
this include: (a) the possibility of a complex Higgs branch topology leading to
extra states, and (b) the fact that when there is a closed loop the Coulomb
branch constraint equation (\ref{coulombDterm}) has a continuous family of
solutions in which the centers approach each other ever more closely.\footnote{An
easy way to see this, is that the left hand side of all constraint equations
will  always involve at least two terms with opposite sign and so always have a
solution for vanishing separation.}    The latter fact suggests that for any
$g_s$ there will be some states whose wavefunctions have substantial support on
the Higgs branch.      We are proposing a different perspective.   In our view,
the transverse growth of the size of a bound state as the system flows from
Higgs branch to Coulomb branch to a closed string description is responsible
for the formation of a complex macroscopic structure with an effective
description as a black hole.  In this perspective most microstates should enjoy
such a flow and the usual solution with a horizon is simply the effective
long-wavelength description of many complex, spatially extended microscopic
bound states \cite{mathur1,mathur2,bw2,us,BKS,LLM,BdBJS,integ,inforecovery}.

\section{Discussion}

We make two proposals in this article.  First, we  suggest that every supersymmetric four dimensional black hole of finite area can be split up into microstates made of primitive 1/2-BPS ``atoms''.   The non-locality of the charges of these atoms binds these solutions together.   Secondly, we propose that at very weak coupling these states appear as bound D-branes, but that as the coupling grows the bound state grows a transverse size leading to a topologically complex spacetime with an effective description as a black hole.  At strong coupling the states form a ``foam'' in M-theory.    To provide evidence for our proposal we constructed a large class of smooth, horizon-free supergravity solutions  with the charges of four dimensional black holes, and demonstrated a scaling relation that takes them, as $g_s \to 0$, from a foam in M-theory, to multi-centered solutions in four-dimensional supergravity, to states in the D-brane gauge theory, first in the Coulomb branch and then in the Higgs branch.    Our gauge theory analysis extensively used the results of Denef \cite{denefhall}, who explicitly studies the flow of quantum mechanical wavefunctions from Coulomb to Higgs branch in some examples.   We are also proposing that the reverse of this process, the flow of states from the Higgs branch into the Coulomb branch and then into a closed string description, is responsible for a transverse growth in the size of D-brane bound states as the string coupling increases, and that this is the link between the D-brane and ``spacetime foam'' pictures of black hole microstates.

\begin{figure}
\centering \hspace{0.0in} \includegraphics[width=1.0\textwidth]{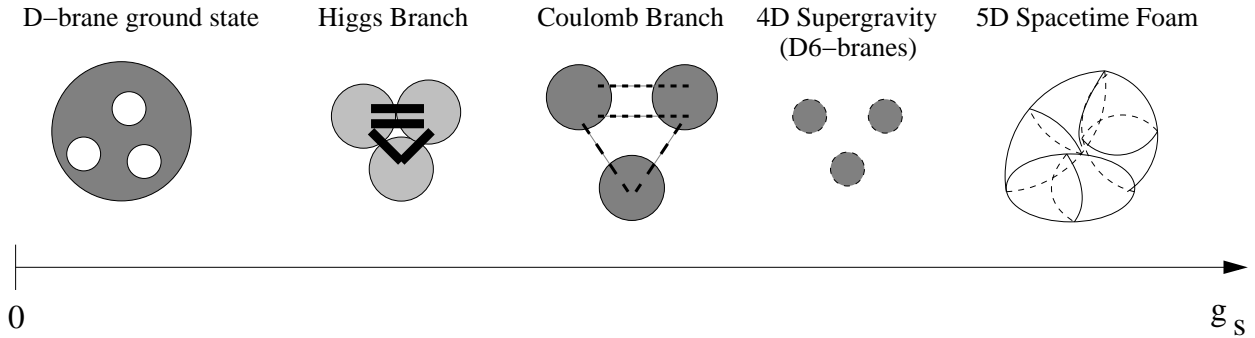} 
\caption{The different phases as we increase $g_s$.}
\end{figure}

To prove our proposals there are several further steps that must be taken
\begin{enumerate}
\item We must demonstrate that there are enough microstates
constructed from 1/2-BPS ``atoms'' to account for the known entropy of
the black hole carrying the total charge of the system.
\item We should show that the typical microstate at finite string
coupling has a complex structure out to the horizon scale, but that
the detailed microstructure is inaccessible to a conventional
semiclassical observer.
\item We must complete our understanding of the relation of the
Coulomb and Higgs branches of quiver gauge theories, in particular
whether spacetime constraints such as triangle inequalities are
realized in the Higgs branch also.
\item We must understand the role of the superpotentials that appear
in quivers with closed loops in determining the structure of the Higgs
branch moduli space, and whether and how this affects the flow of
states between these branches as the coupling changes.
\end{enumerate}
While these are challenging problems, solving them is important for  understanding the quantum mechanical states underlying classical spacetimes.

\paragraph{Acknowledgments: } We thank
Jan de Boer, Frederick Denef, Michael Douglas, Sergio Ferrara, Davide Gaiotto,
Ori Ganor, Renata Kallosh, Boris Pioline and Joan Sim\'{o}n for useful
discussions. This work was supported by the DOE grant DE-FG02-95ER40893
(VB,TL), the NSF grant PHY-0331728 (VB,TL), and a dissertation fellowship from
the University of Pennsylvania (TL). The work of EG was supported by the US
Department of Energy under contracts DE-AC03-76SF00098 and DE-FG03-91ER-40676
and by the National Science Foundation under grant PHY-00-98840.  VB thanks the
Weizmann Institute for hospitality while this paper was completed.  
EG and VB would like to dedicate this paper to the memory of John Brodie and Andrew Chamblin.

\appendix
\section{The central charge for $\alpha=0$} \Label{app-cent}
The calculation of the $\CN =2$ central charge is outlined in
\cite{denef0brane} and is given by
\be
\label{Zdef}
Z_p = F_\mu p_p^\mu - X^\mu q^p_\mu , \Label{centexpr}
\ee
where $X^\mu$ are the projective coordinates for the four-dimensional scalars
\beq
\label{Xdef}
X^0 &=& {-1\over \sqrt{J_4}} {\partial J_4 \over \partial K^0} - i M_0 \to i \ \ \rm{as} \ \rho \to \infty, \nonumber \\
X^i &=& {-1\over \sqrt{J_4}} {\partial J_4 \over \partial M_j} + i K^j \to (-1 + i \delta K^j) .
\eeq
$F_\mu =\partial F(X) / \partial X^\mu$ is the derivative of the cubic prepotential\footnote{$F$ takes 
this simple form as a result of compactifying on $T^6$ and only turning on the eight charges of the STU model. 
A more general case will alter this expression in a straightforward manner.}
\be
F(X) = -{X^1 X^2 X^2 \over X^0} .
\ee
In the conventions of \cite{denef0brane} the magnetic and electric charges for the $p$th brane are given by
\footnote{The reader will notice that our conventions exchange some electric and magnetic pairs by taking a 
series of Hodge duals. }
\beq
p^0 &=& {L \over 2} n_p , \ \ \ \ \ p^j = {1 \over 2} d^j_p \nonumber \\
q_0 &=& {1 \over 2L^2} {d^1_p d^2_p d^3_p \over n_p^2} , 
\ \ \ q_i = - {1 \over 2L} s_{ijk} {d^j_p d^k_p \over 2n_p} .
\eeq
Putting this all together we find
\be
Z_p = {L \over 2} n_p - {1 \over 2L} \sum_{i,j,k} {s_{ijk} \over 2} n_p 
\lambda^j_p \lambda^k_p +{i \over 2} \Psi_p \, .
\ee
The mass of the $p$th brane is given by
\be
m_p = { |Z_p| \over 4 G_4} . \Label{branemass}
\ee
Note that the total central charge is $Z = \sum_p Z_p$, and that
\be
\label{impart}
Im[Z] = {1\over 2} \sum_p \Psi_p = 0
\ee
because of the integrability constraint and the fact that we have set
$\alpha$ in (\ref{deltaK}) to zero for simplicity.

\section{Non-zero $\alpha$}

In general, for non-zero values of $\alpha$ the situation is more
complicated as the reduction from five to four dimensions is now along a fiber
of a slightly different magnitude (a similar situation arises in \cite{EEMR}). 

Let us consider how $\alpha \ne 0$ affects the phase of the central charge $Z$ as defined above.
Note first that the expression $Im[Z_p] = {1\over 2}\Psi_p = 2\!<\!\!\Gamma_p,\Gamma_\infty\!\!>$ 
is a trivial consequence of equations (\ref{Zdef}), (\ref{Xdef}) and the explicit form for the $F_\mu$'s:
\be
F_0 = {+1\over \sqrt{J_4}} {\partial J_4 \over \partial M_0} + i K_0 , \qquad
F_i = {-1\over \sqrt{J_4}} {\partial J_4 \over \partial K^j} + i K_j. 
\ee
This implies that eq. (\ref{impart}) still holds, hence the central charge is real and $\alpha$ 
cannot be it's phase!
It turns out that the phase $\alpha$ of the central charge used in \cite{denef1,denef3,denefhall} 
is defined in a different gauge, where a Kahler transformation
has been used to fix the asymptotic value of $X^0\to i$.  
As we will demonstrate, for non-zero $\alpha$ our asymptotic value
for $X_0$ is $ie^{-i\alpha}$.  Rotating this back to 
the Denef etal.'s gauge implies that the central charge picks up an overall phase of 
$\alpha$.  Hence $\alpha$ is the phase of of the central charge in Denef etal.'s gauge.  
Note that the expression $\Psi_p = 
4\!<\!\!\Gamma_p,\Gamma_\infty\!\!>$ is gauge invariant.  We will demonstrate that even
for non-zero $\alpha$ the FI terms $\theta_p$ are 
stil exactly equal to $(8G_4)^{-1}\Psi_p$, as expected.

\subsection{Redefining the harmonics for $\alpha\ne 0$}

We start by observing that for non-zero $\alpha$ as defined in eq. (\ref{deltaK}) the asymptotic
value of $k_0$ becomes ${L\over 2} \sin\alpha$.  This implies that if we left our definitions for the
harmonic functions $M^\mu, K_\mu$ unchanged, the asymptotic value for $J_4$ would now be $\cos^2\alpha < 1$.
To remedy this situation we need to adjust the reduction of our five-dimensional solution to four dimensions 
as follows. First we recognize that the IIA coupling constant is now $g_sl_s = L\cos\alpha = R$, but we still 
define the new radial coordinate $\rho = 2r/L$.  The time coordinate now also needs to be rescaled 
$t^{(4D)} = \sec\alpha\, t^{(5D)}$.  The new harmonic functions are:
\be
    M_0 = - \cos\alpha\, H L^2/4,\;\; K^0 = \sec^2\alpha\, 4K/L, \;\; K^i = L h^i,\;\; M_i = \sec\alpha\, M_i^{5D},
\Label{dimharm1} 
\ee 
with similar scalings for $\Gamma$.
With these definitions, $J_4(x) = I_4({\cal H}(x)) \to 1$ at $\infty$.  The asymptotic value for
$X^0$ is now $X^0 \to \sin\alpha + i \cos\alpha = ie^{-i\alpha}$ as previously advertised.

\subsection{Checking the FI term}

  In the gauge of Denef etal., it also possible to write down the
individual contribution to the central charge from each center as: \be
Z_p = {R\over 2} n_p \prod_i (1 + i B^i_p) = 4G_4\, m_p\,
e^{i\alpha_p}, \ee where the $B^i_p = \({d_p^i\over R n_p}- {\delta
K^i\over \cos\alpha} + {\sin\alpha\over\cos\alpha}\)$ are the
normalized (F-B) terms on each $D6$-brane.  This allows to quickly
check that we have the right definition for the $FI$ term: \beq
(8G_4)\, \theta_p &=& 2|Z_p|\sin(\alpha_p - \alpha) \\ &=&
2|Z_p|\sin\alpha_p\cos\alpha - 2|Z_p|\cos\alpha_p\sin\alpha \nonumber
\\ &=& R n_p\, Re\[\prod_i (1 + i B^i_p)\]\cos\alpha - R n_p\,
Im\[\prod_i (1 + i B^i_p)\]\sin\alpha \nonumber \\ &=& R n_p \[ \sum_i
B^i_p - \prod_i B^i_p \] \cos\alpha - R n_p \[1 + \sum_i {s_{ijk}\over
2}\, B^j_pB^k_p\]\sin\alpha \nonumber\\ &=& {n_p\over \cos\alpha}\(
\sum_i\lambda^i_p - {1\over L^2} \prod_i \lambda^i_p + 2L\sin\alpha \)
\nonumber \\ &=& \Psi_p^{4D} = 4 <\Gamma_p,\Gamma_\infty> =
\Psi_p^{5D}/\cos\alpha.  \eeq Here we have differentiated $\Psi_p^{4D}
= 4 <\Gamma_p,\Gamma_\infty>$ which is defined canonically in $D=4$
from the renormalized charge vectors $\Gamma_p$ from 5D inspired $\Psi_p^{5D}$ 
defined by simply adding a correction term $(2Ln_p\sin\alpha)$ to the
right-hand side of eq.(\ref{extra}).


\providecommand{\href}[2]{#2}\begingroup\raggedright\endgroup

\end{document}